\documentclass{article}

\usepackage{microtype}
\usepackage{graphicx}
\usepackage{subfigure}
\usepackage{booktabs} 
\usepackage{amsfonts,amssymb,amsthm,amsmath}
\usepackage{nicefrac}  
\usepackage{natbib}
\usepackage{hyperref}
\usepackage{tikz}

\usepackage{fullpage}
\usepackage{xspace}

\usepackage{algorithm}
\usepackage{algorithmic}

\usepackage{authblk}

\newcommand{\ourDataset}{\textsc{Cumulo}\xspace}
\newcommand{\myparagraph}[1]{\vspace{1mm}\noindent{\bf #1}}

\newcommand{\Xb}{\mathbf{X}}
\newcommand{\Sb}{\mathbf{S}}
\newcommand{\Lb}{\mathbf{L}}
\newcommand{\Cb}{\mathbf{C}}
\newcommand{\Vb}{\mathbf{V}}
\newcommand{\yb}{\mathbf{y}}
\newcommand{\zb}{\mathbf{z}}

\newcommand{\bb}[1]{{\mathbf{#1}}}

\newcommand{\Xc}{\mathcal{X}}
\newcommand{\Yc}{\mathcal{Y}}
\newcommand{\Nc}{\mathcal{N}}
\newcommand{\Lc}{\mathcal{L}}
\newcommand{\Uc}{\mathcal{U}}

\title{Cumulo: A Dataset for Learning Cloud Classes\footnote{This work was performed during the research accelerator Frontier Development Lab Europe.}}

\date{}
\begin{document}

\author[1]{Valentina Zantedeschi\thanks{vzantedeschi@gmail.com}}
\author[2]{Fabrizio Falasca\thanks{fabrifalasca@gatech.edu}}
\author[3]{Alyson Douglas}
\author[4]{Richard Strange}
\author[5,6]{\\Matt J. Kusner\thanks{mkusner@turing.ac.uk}}
\author[4]{Duncan Watson-Parris\thanks{duncan.watson-parris@physics.ox.ac.uk}}

\affil[1]{GE Global Research, San Ramon, USA}
\affil[2]{Georgia Institute of Technology, Atlanta, USA}
\affil[3]{University of Wisconsin, Madison, USA}
\affil[4]{University of Oxford, Oxford, UK}
\affil[5]{The Alan Turing Institute, London, UK}
\affil[6]{University College London, London, UK}

\maketitle

\begin{abstract}
One of the greatest sources of uncertainty in future climate projections comes from limitations in modelling clouds and in understanding how different cloud types interact with the climate system.
A key first step in reducing this uncertainty is to accurately classify cloud types at high spatial and temporal resolution.
In this paper, we introduce \ourDataset,
a benchmark dataset for training and evaluating global cloud classification models. It consists of one year of 1km resolution MODIS  hyperspectral imagery merged with  pixel-width `tracks' of CloudSat cloud labels. 
Bringing these complementary datasets together is a crucial first step, enabling the Machine-Learning community to develop innovative new techniques which could greatly benefit the Climate community.
To showcase \ourDataset, we provide baseline performance analysis using an invertible flow generative model (IResNet), which further allows us to discover new sub-classes for a given cloud class by exploring the latent space. To compare methods, we introduce a set of evaluation criteria, to identify models that are not only accurate, but also physically-realistic.
\ourDataset~can be download from \url{https://www.dropbox.com/sh/i3s9q2v2jjyk2it/AACxXnXfMF5wuIqLXqH4NJOra?dl=0}.
\end{abstract}

\section{Cloud Classification is key for modelling Climate Change}
Clouds play a crucial role in the climate system. They are the source of all precipitation and have a significant impact on the Earth's radiative budget. Crucially, as any changes in clouds impact the environment; these changes feedback on cloud formation and behaviour. These feedbacks are a primary source of uncertainty for climate model projections \citep{knutti, Rotstayn, stephens}, as there is a limited understanding of the mechanisms and relationships between clouds, climate and global circulation \citep{bony_Nature}. It is for example unclear how warmer sea surface temperature will affect clouds and convective organization \citep{bony_Nature}, or trigger possible climate transitions by cloud breakup \citep{schneider}. Comprehensive use of the vast observational data available is crucial to improve our understanding of these processes, and their representations in climate models.

Clouds can form and develop through several different pathways, depending on their environment and the convective energy available. It is common to categorise clouds into different types based on their properties to better analyse them. The International Satellite Cloud Climatology Project (ISCCP) dataset \citep{Rossow1991} provides a global classification of clouds at a 10km resolution, based on a network of geostationary meteorological satellites. Satellite based observations of clouds can be made using either passive imagery or active radar instruments. While high-resolution hyperspectral imagery is available from both polar orbiting, and geostationary satellites, providing excellent coverage at high temporal resolution, specific cloud properties (such as their exact height and droplet size distribution) must be inferred indirectly. The ISCCP classification relies on a simple assessment of the relationship between the clouds' inferred height and optical thickness \citep{Rossow1991}. Conversely, the CloudSat cloud radar does provide direct measurements of clouds and their properties. This comes with a drawback, as it operates with a narrow swath and and repeat a cycle every 16 days (no global coverage at 1km resolution is present even after 16 days).

To overcome these limitations, in this paper we introduce \ourDataset, a new dataset which combines the global 1km-resolution imagery of the Moderate Resolution Imaging Spectroradiometer (MODIS) with the accurately measured properties of the CloudSat products. It contains one year of 1354 x 2030 pixel hyperspectral images from MODIS combined with pixel-width `tracks' of cloud labels from Cloudsat, corresponding to the eight World Meteorological Organization (WMO) genera (Fig.~\ref{fig:dataset}). While both datasets are publicly available, the extraction, cleaning and alignment of the data required specialist domain knowledge and extensive compute resources.

We apply a deep generative model architecture on one month of \ourDataset, and present, for the first time to our knowledge, global high resolution spatiotemporal cloud classification derived from a combination of active and passive satellite sensors.
We show that our results are physically reasonable in terms of locations of occurrences of the given classes and liquid water path distributions. 

\myparagraph{Related Work.}
\citet{muhlbauer2014climatology} classify MODIS into three types of mesoscale cellular convection using a 3-layer neural network. While these classifications are global, they only describe a particular climatology. 
\citet{zhang2019development} classify the geostationary Himawari-8 satellite into WMO cloud classes (from Cloudsat) using a random forest. Alternatively, our dataset provides global coverage (versus East Asia and Western Pacific for Himawari-8). 
\citet{rasp2019combining} crowd-sourced human-level classifications of shallow trade clouds into four types: `sugar', `flower', `fish', `gravel'. They  evaluate an object detection method \cite{lin2017focal} and a semantic segmentation method \cite{ronneberger2015u} to classify clouds into these types. 
Similar to \cite{muhlbauer2014climatology}, this work only aims at a small aspect of cloud variability. 

\section{\ourDataset: A global dataset for cloud classification} \label{Dataset_Section}

The proposed dataset contains 105,120 geolocated and hyperspectral images and provides a combination of channels from different sources (see Table~\ref{tab:channels}): 
the selected radiance channels from MODIS AQUA Calibrated Radiances fully capture the physical properties needed for cloud classification and are meant to be used for training; MODIS AQUA Cloud Product channels are retrieved features describing cloud physical properties useful for validation; MODIS AQUA Cloud Mask detects the presence of a cloud; 2B-CLDCLASS-LIDAR provides the types of clouds spotted at different heights along the track of the satellites and additional cloud characteristics.

Possible cloud types, corresponding to the eight WMO genera, are stratus (St), stratocumulus (Sc), cumulus (Cu, including cumulus congestus), nimbostratus (Ns), altocumulus (Ac), altostratus (As), deep convective (cumulonimbus, Dc), and high (Ci, cirrus and cirrostratus).
Refer to Table~\ref{tab:classes} for a synthetic description of the classes.
2B-CLDCLASS-LIDAR variables are defined on up to 10 different vertical layers. 
Each layer corresponds to an identified cloud cluster at the given time and location.
Because the type and the number of identified clouds inevitably vary over space and time, layers are not predefined intervals of fixed size over the altitude, but their number and thickness vary over the pixels. 
For instance, in clear sky conditions no cloud layers are reported, while multiple layers are listed whenever low and high clouds coexist at different altitudes.
It is worth noticing that cloud layers are considered distinct if their hydrometeor-free separation is of at least ~480 m.
Cloud layer base and top variables provide the information for inferring the height and thickness of the identified layers, cloud type quality is the confidence in the labels based on fuzzy-logic classification, and precipitation flag indicates whether any of the cloud layers produce precipitation.

Finally, notice that some channels are available only at daytime, because they rely directly or indirectly on daylight radiances.
In general, missing values due to artefacts were filled with the nearest (in time and space) available values.

\begin{table}[h!]
  \caption{Channel descriptions}
  \vspace{-0.5ex}
  \label{tab:channels}
  \centering
  \begin{scriptsize}
  \begin{tabular}{llcll}
    \toprule
    Source & Name/Description  & Index & Primary Use & Availability \\
    \midrule
    MODIS  & shortwave visible (red) & 1  & land/shadow/cloud/aerosols boundaries & daytime \\
      & shortwave near infrared & 2  & land/shadow/cloud/aerosols boundaries & daytime \\
      & longwave thermal-infrared & 20-23  & surface/cloud temperature & always\\
      & shortwave near infrared  & 26 & Cirrus clouds water vapor & daytime \\
      & longwave thermal-infrared  & 27 & water vapor & always\\
      & longwave thermal-infrared  & 29 & cloud properties & always\\
      & longwave thermal-infrared  & 33-36 & cloud top altitude & always\\
    \midrule
    MODIS Cloud Product & liquid water path  & & physical validation & always\\
    & cloud optical thickness & & physical validation & daytime \\
    & cloud effective radius  & & physical validation & daytime\\
    & cloud particle phase  & & physical validation & daytime\\
    & cloud top pressure  & & physical validation & always\\
    & cloud top height  & & physical validation & always\\
    & cloud top temperature & & physical validation & always \\
    & cloud effective emissivity  & & physical validation & always\\
    & surface temperature && physical validation & always \\
    \midrule
    MODIS Cloud Mask & cloud mask & & cloud detection & always \\
    \midrule
    2B-CLDCLASS-LIDAR & cloud layer type & & cloud labels & always \\
    & & & & (limited coverage, 3D) \\
     & cloud layer base & & cloud base altitude & always \\
    & & & & (limited coverage, 3D) \\
     & cloud layer top & & cloud top altitude & always \\
    & & & & (limited coverage, 3D) \\
     & cloud type quality & & label quality & always \\
    & & & & (limited coverage, 3D) \\
    & precipitation flag & & cloud properties & always \\
    & & & & (limited coverage) \\
    \bottomrule
  \end{tabular}
  \end{scriptsize}
\end{table}

\begin{table}[h!]
  \caption{Class descriptions}
  \vspace{-0.5ex}
  \label{tab:classes}
  \centering
  \begin{scriptsize}
  \begin{tabular}{clccccc}
    \toprule
    Index & WMO Name & \multicolumn{4}{c}{Characteristics} & Proportion \\
    && thickness & base height & liquid water path & rain & \\
    \midrule
    0 & Cirrus and cirrostratus (Ci) & moderate & $>$ 7.0 km & 0. & none & $30.68\%$ \\
    1 & Altostratus (As) & moderate & 2.0-7.0 km & $\sim$ 0. & none &  $16.02\%$ \\
    2 & Altocumulus (Ac) & shallow/moderate & 2.0-7.0 km & $>$ 0. & virga possible & $9.53\%$ \\
    3 & Stratus (St) & shallow  & 0-2.0 km & $>$ 0. & none/slight & $1.84\%$ \\
    4 & Stratocumulus (Sc) & shallow  & 0.-2.0 km & $>$ 0. & drizzle/snow possible & $27.53\%$ \\
    5 & Cumulus (Cu) & shallow/moderate & 0-3.0 km & $>$ 0. & drizzle/snow possible & $6.02\%$ \\
    6 & Nimbostratus (Ns) & thick & 0-4.0 km & $>$ 0. & prolonged rain/snow & $7.40\%$ \\
    7 & Deep Convection (Dc) & thick & 0-3.0 km & $>$ 0. & intense rain/hail & $0.96\%$ \\
    \bottomrule
  \end{tabular}
  \end{scriptsize}
\end{table}

\begin{figure}[h!]
\centering
\subfigure[All swaths of one day, projected on the globe, with their label masks. Notice that swaths can overlap and that, for visualisation purposes, label tracks are magnified.]{\label{fig:day} \includegraphics[width=0.7\textwidth]{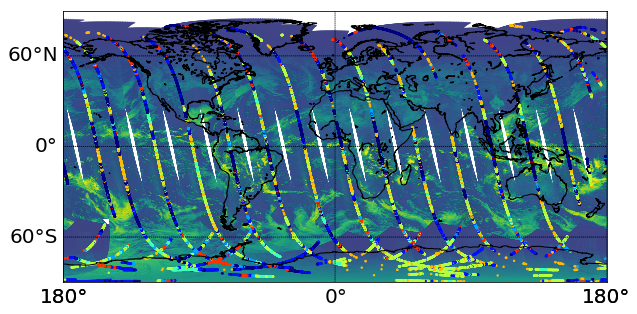}}
\subfigure[The visible band (left) and the cloud mask (right) of a sample swath with its overlying label mask.]{\label{fig:swath}
\includegraphics[width=0.7\textwidth]{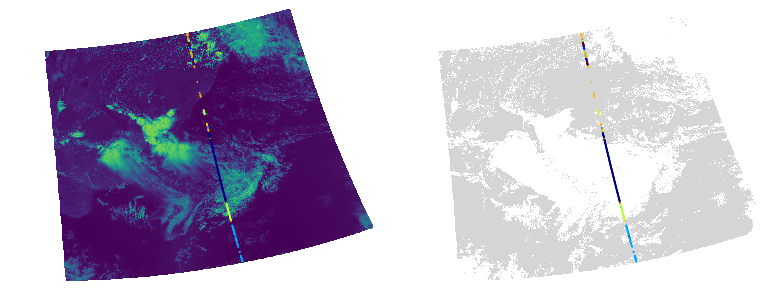}}
\caption{Visualization of \ourDataset's data coverage.}
\label{fig:dataset}
\end{figure}

More precisely, each satellite image (swath) is acquired at a given time $t$ (one swath every five minutes) and at a given location $l$ (each pixel is associated with a latitude-longitude pair): $\{ \Sb^{t,l} \}_{t=1,\ldots,T ; l=1,\ldots,L}$.
In the following, we denote by
\begin{itemize}
    \item $\Xb^{t,l} \in \mathbb{R}^{2030 \times 1354 \times 13}$ the thirteen training channels coming from MODIS AQUA Calibrated Radiances~\footnote{\url{https://modis.gsfc.nasa.gov/data/dataprod/mod02.php}};
    \item $\Vb^{t,l} \in \mathbb{R}^{2030 \times 1354 \times 9}$ the eight validation channels coming from MODIS AQUA Cloud Product~\footnote{\url{https://modis.gsfc.nasa.gov/data/dataprod/mod06.php}} which provide physical and radiative cloud properties obtained by combining infrared emission and solar reflectance techniques applied on MODIS original bands;
    \item $\Cb^{t,l} \in \mathbb{R}^{2030 \times 1354}$ the cloud mask derived from MODIS Cloud Mask~\footnote{\url{https://modis.gsfc.nasa.gov/data/dataprod/mod35.php}}, marking as $1$ the certainly cloudy pixels and as $0$ any other pixel;
    \item $\Lb^{t,l} \in \mathbb{R}^{2030 \times 1354}$ the overlay label mask derived from 2B-CLDCLASS-LIDAR~\footnote{\url{http://www.cloudsat.cira.colostate.edu/data-products/level-2b/2b-cldclass-lidar}}, indicating for each pixel the most frequent type of clouds that were identified at different heights.
\end{itemize}  
Figure~\ref{fig:dataset} shows a snapshot of \ourDataset~for one day for the whole globe (Fig.~\ref{fig:dataset}a) and for one swath (Fig.~\ref{fig:dataset}b), along with its MODIS cloud mask (Fig.~\ref{fig:dataset}c). 

Overall, \ourDataset~provides a comprehensive set of features both for identifying cloud types and for validating any finding, along with ready-to-use and accurate cloud annotations at high spatial resolution.
From a Machine Learning perspective, \ourDataset~presents several challenges: supervision is available only for 1 every 1354 pixels (weakly-labelled data), pixels can be annotated with multiple types of clouds (multi-labelled data), and many cloud classes, such as deep convective clouds, are underrepresented (class imbalance).

\newpage

\section{Applying Cloud Classification globally} \label{ML_Section}

In this section, we provide baseline performance analysis of one of the tasks that can be performed using \ourDataset:  semi-supervised classification of clouds at a global and daily scale. 

\subsection{Methodology} Given the small number of labels, we find that performing this experiment at a pixel level allows us to achieve better classification performance than applying common semantic segmentation models (such as~\cite{papandreou1502weakly,li2018weakly,ronneberger2015u}) that necessitate full label masks or annotations at the instance or image level.
We consider tiles of 3x3 pixels extracted from the first month of data (January 2008) and predict a label for each tile. We use the most frequent cloud type identified within each tile as a target, when annotations are available.
We have gathered two sets of tiles: a set of labelled tiles sampled around any annotated pixel on the satellite's track; and a sample of unlabelled tiles randomly selected from the non-annotated regions.
We have made use of the available cloud mask for restraining the classification to tiles that had a high probability of cloud cover.
We have trained a hybrid Invertible Residual Network~\citep{nalisnick2019hybrid} which allows us to (i) harness both labelled and unlabelled sets and (ii) learn a representation where the class distributions can be further subdivided into fine-grained classes.
Indeed, cloud types are not limited to the well-studied WMO genera: 
being able to identify more species of clouds is an open-question in the cloud community.
Our model combines a deep generative flownet with a linear classifier, which is simultaneously trained by maximizing the joint log-likelihood over the tiles and their labels~\footnote{label entropy is minimised instead of the cross-entropy whenever supervision is not available.}.
Classes are weighted in the objective to encourage a better classification of under-represented classes.

\subsection{Formal problem setting} \label{ap:formal}
We segment each tensor $\Xb^{t,l}$ into non-overlapping tiles of size 3x3 pixels, e.g.
    \begin{align}
        \Xb^{t,l} = \begin{bmatrix}
        \Xb_{(0,0)}^{t,l} & \Xb_{(0,1)}^{t,l} & \cdots & \Xb_{(0,W)}^{t,l} \\
        \Xb_{(1,0)}^{t,l} & \Xb_{(1,1)}^{t,l} & \cdots & \Xb_{(1,W)}^{t,l} \\
        \vdots & \vdots & \vdots & \vdots \\
        \Xb_{(H,0)}^{t,l} & \Xb_{(H,1)}^{t,l} & \cdots & \Xb_{(H,W)}^{t,l}
        \end{bmatrix}, 
    \end{align}
and aim at learning a mapping $f: \Xc = \mathbb{R}^{3, 3, 13} \rightarrow \Yc = \{-1,\dots, 7\}$ from any tile $\Xb_{(i,j)}^{t,l}$ to a class label $\yb_{(i,j)}^{t,l}$.
As target values, we retain the most frequent cloud type occurring within the label mask $\Lb_{(i,j)}^{t,l}$ associated to a tile.

For learning, we recall that we make use of two sets of tiles of equal size $K$: a set of labeled tiles $\Lc = \{(x_k, y_k) \in \Xc \times \Yc \}_{k=1}^{K}$, sampled around any annotated pixel on the satellite's track, and a sample of unlabeled tiles $\Uc = \{x_k \in \Xc \}_{k=1}^{K}$, randomly selected from the non-annotated regions.
We deploy a hybrid Invertible Residual Network~\citep{nalisnick2019hybrid}, parameterised by $\theta$, which allows us to learn a latent representation for modelling the true distribution of the data $p_\theta(\Xb)$ by the decomposition
\begin{align}
p_\theta(\Xb) = \sum_{\zb} p_\theta(\Xb | \zb)p(\zb)
\end{align}
where $\zb \in \mathbb{R}^{2030, 1354}$ is the latent representation of $\Xb$ and $p(\zb) \sim \Nc(0, 1)$ its prior distribution.


The peculiarity of flownets, such as the hybrid IResNet, is that they consist of a series of mappings $f = f_1 \circ f_2 \circ \cdots \circ f_D$ that are invertible for any input $x \in \Xc$: $ \forall i=1,\dots,D ; f_i(x)^{-1} = x$. 
Crucially, there is a direct and invertible relationship between each image $\Xb$ and each latent point $\zb$:
\begin{align}
\Xb \longleftrightarrow^{f_1} \bb{h}_1 \longleftrightarrow^{f_2} \bb{h}_2 \cdots \longleftrightarrow^{f_D} \zb 
\end{align}

Thanks to these constraints on the architecture of the network, it is possible to optimise $\theta$ by maximising the joint likelihood of the swaths $\Xb$ and their target label masks $\yb$, rewritten as:
\begin{align}
    p_\theta(\yb, \Xb) &= p(\yb \mid \Xb) p(\Xb) \\
    &= p(\yb \mid \zb) p(\zb) \Big| \det \Big( \frac{d\zb}{d\Xb} \Big) \Big| \\
    &= p(\yb \mid \zb) p(\zb) \prod_{i=1}^D \Big| \det \Big( \frac{d\bb{h}_i}{d\bb{h}_{i-1}} \Big) \Big| \label{eq:likel}
\end{align}
where the last two terms are given by the change of variable formula. 
As supervision is not provided for all the tiles, $p(\yb \mid \zb)$ in Eq.~\eqref{eq:likel} can be estimated only for tiles from the labeled set $\Lc = \{(x_k, y_k) \}_{k=1}^{K}$. 
For the tiles of the unlabeled set $\Uc = \{x_k \}_{k=1}^{K}$, the label entropy is minimised instead, to promote sharp predictions.
The overall objective function takes the following form, by equivalently maximizing the log likelihood:

\begin{align}
    \max & \sum_{(x_k, y_k) \in \Lc} \left(\log p(y_k | z_k) + \log p(z_k) \right) + \sum_{x_k \in \Uc} \left( \sum_{y \in \Yc}  \log p(y | z_k) + \log p(z_k) \right) + \nonumber \\
    & + \sum_{j=1}^D \log \Big| \det \Big( \frac{d h_j}{d h_{j-1}} \Big) \Big|
\end{align}

\subsection{Results} 
\begin{figure}[h!] 
\centering
\subfigure[Predictions of one day.]{
  \includegraphics[width=0.9\textwidth]{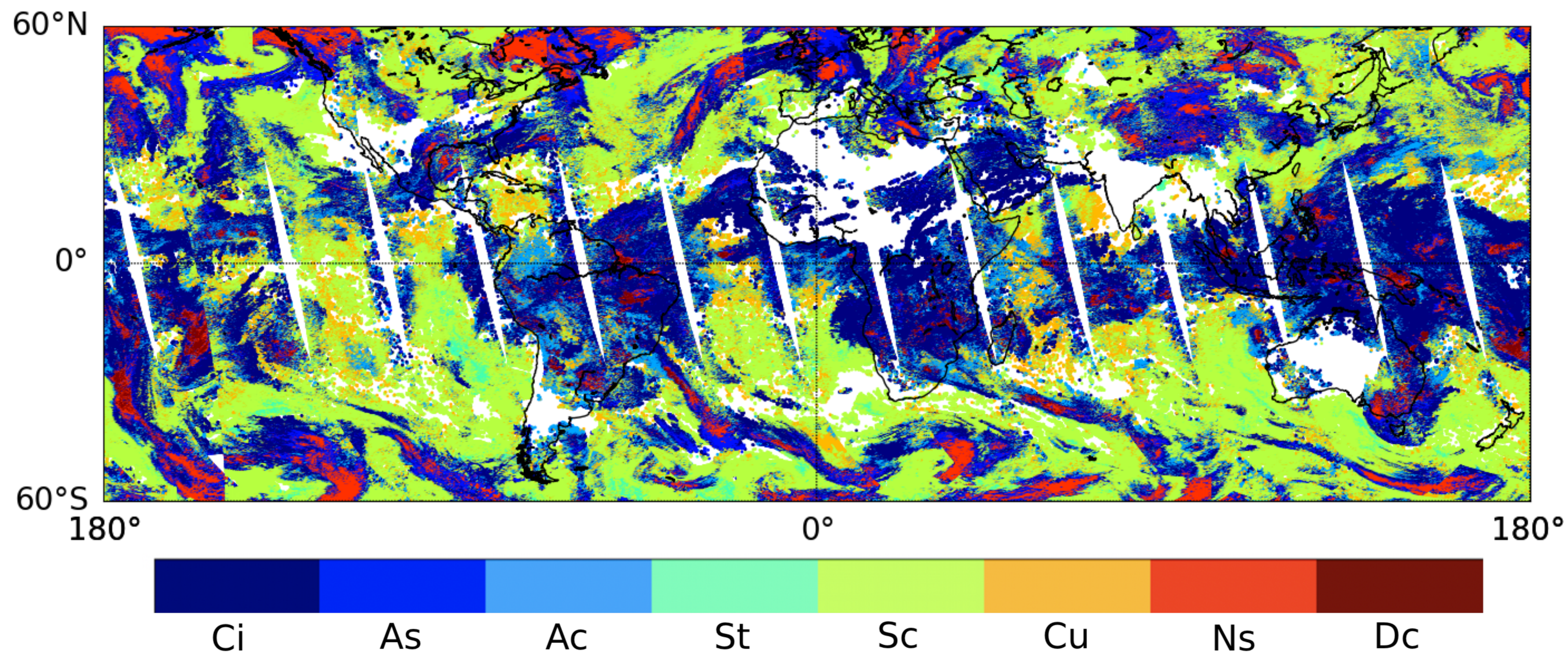}}
  \hfill
\subfigure[Monthly occurrences of Ci clouds.]{
  \includegraphics[width=.31\textwidth]{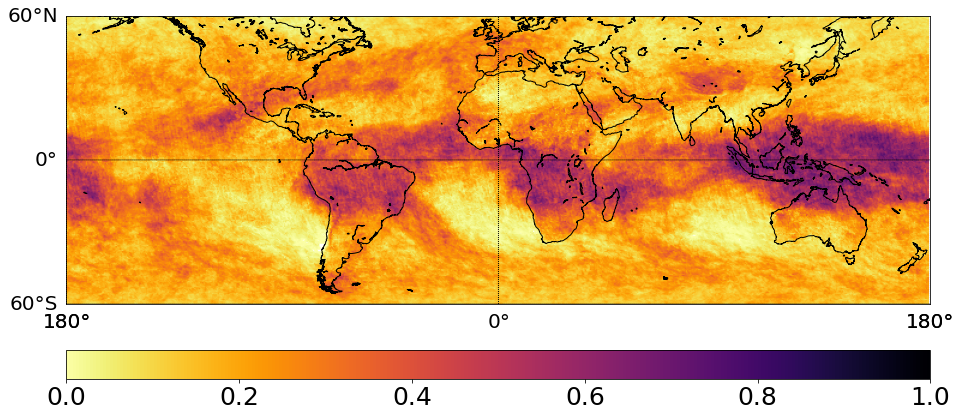}}
\subfigure[Monthly occurrences of Sc clouds.]{
  \includegraphics[width=.31\textwidth]{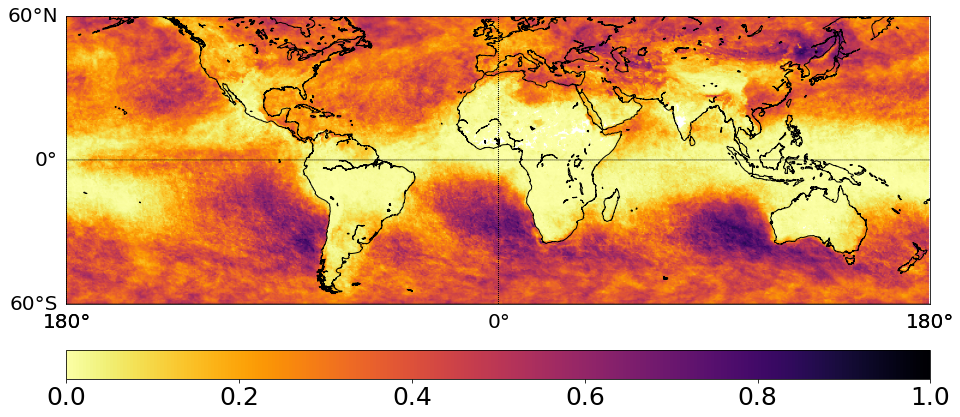}}
\subfigure[Monthly occurrences of Dc clouds.]{
  \includegraphics[width=.31\textwidth]{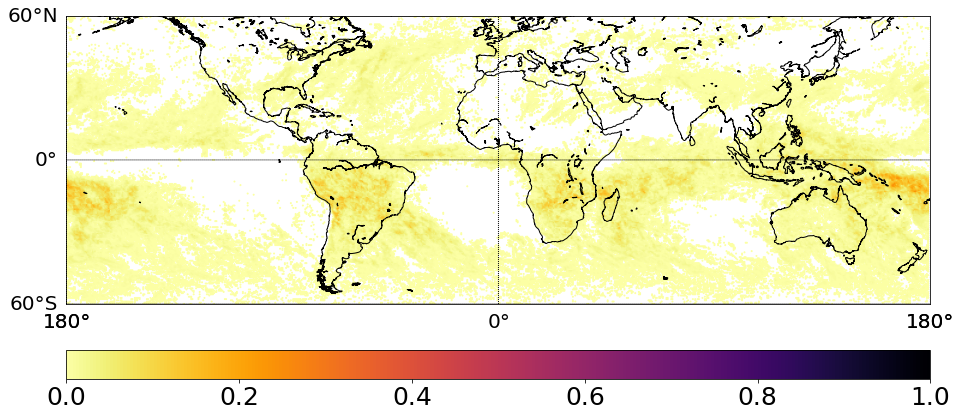} }
\caption{IResNet classification results. Occurrences are computed for January 2008. Occurrences of all classes are shown in Figure~\ref{fig:occurrences}.}
\label{fig:results}
\end{figure}

\begin{figure}[h!]
\subfigure[Ci clouds.]
  {
  \includegraphics[width=.50\textwidth]{figures/occurences-cls-0-logWeights.png}
  }
  \hfill
\subfigure[As clouds.]
  {
  \includegraphics[width=.50\textwidth]{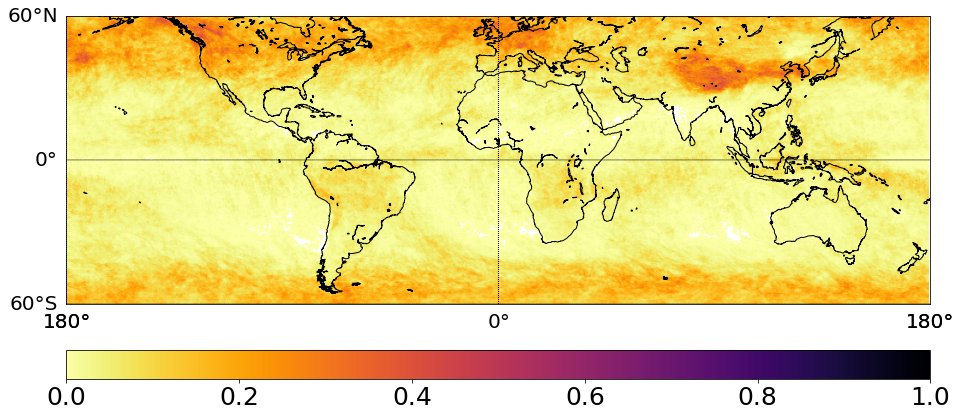}
  }
\subfigure[Ac clouds.]
  {
  \includegraphics[width=.50\textwidth]{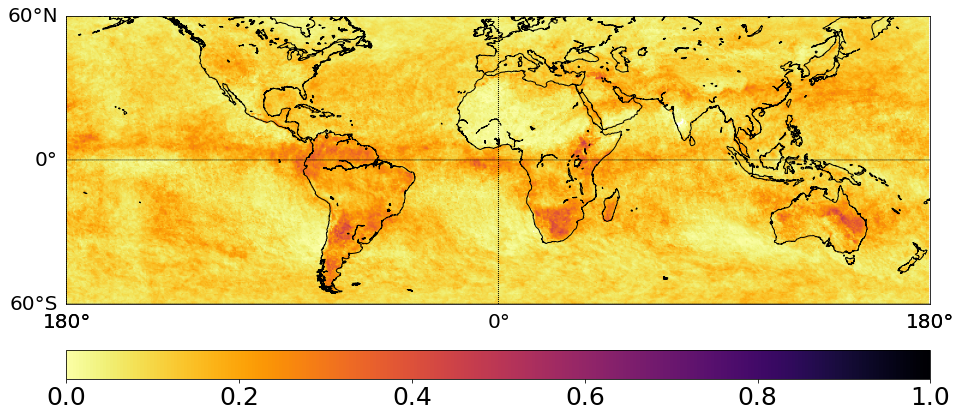}
  }
  \hfill
\subfigure[St clouds.]
  {
  \includegraphics[width=.50\textwidth]{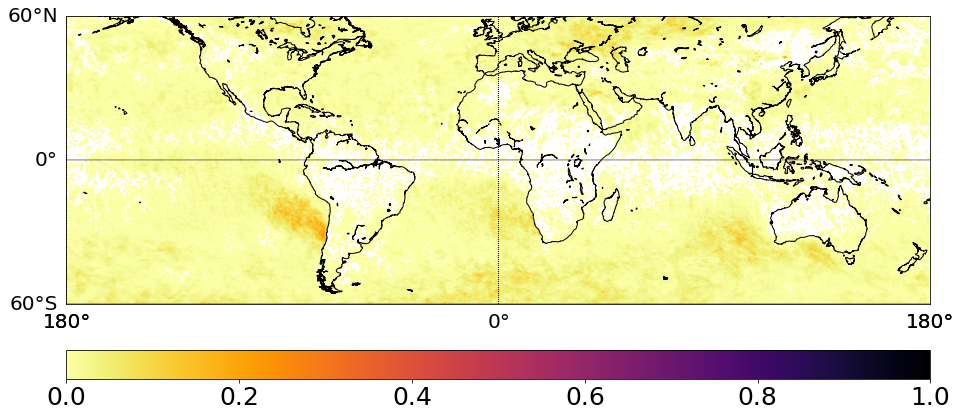}
  }
\subfigure[Sc clouds.]
  {
  \includegraphics[width=.50\textwidth]{figures/occurences-cls-4-logWeights.png}
  }
  \hfill
\subfigure[Cu clouds.]
  {
  \includegraphics[width=.50\textwidth]{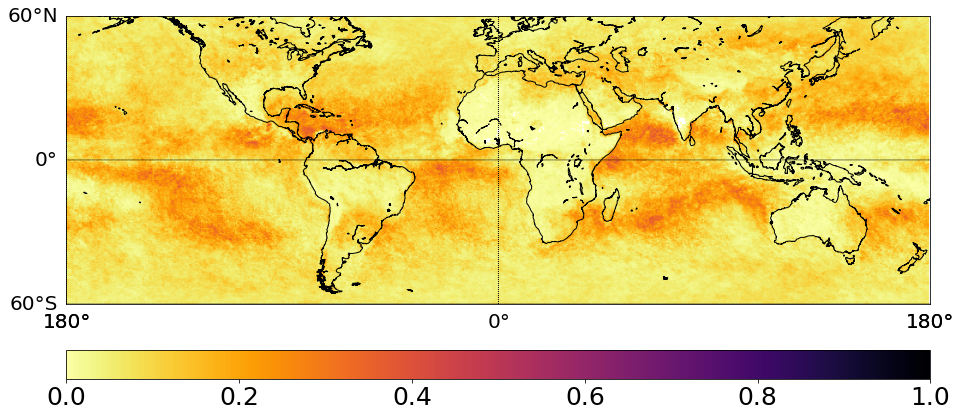}
  }
\subfigure[Ns clouds.]
  {
  \includegraphics[width=.50\textwidth]{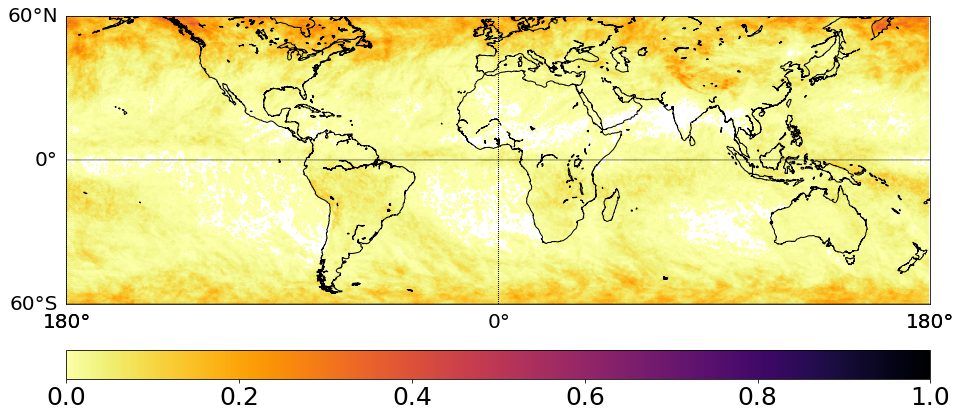}
  }
  \hfill
\subfigure[Dc clouds.]
  {
  \includegraphics[width=.50\textwidth]{figures/occurences-cls-7-logWeights.png}
  }
\caption{Occurrences of the cloud classes predicted by IResNet for January 2008.}
\label{fig:occurrences}
\end{figure}

We randomly split the labelled tiles into training (70\%), validation (10\%) and test (20\%) sets. We report test classification accuracies, F1 score and Intersection over Union index, per class and on average, in Table~\ref{tab:perfs} for the model with the best mean accuracy on the validation set. Figure~\ref{fig:results} shows the predictions obtained over one day of images and the occurrences (gridded at $0.5^{\circ}$ latitude $ \times 0.5^{\circ}$ longitude) of three predicted classes over the month. Predicted classes (Fig.~\ref{fig:results}a) are spatially contiguous across swaths (this is not a constraint of our algorithm). The occurrences appear spatially coherent with Sc clouds occurring mostly over upwelling regions of the major oceans (Fig.~\ref{fig:results}c); Dc clouds are more confined to equatorial regions (Fig.~\ref{fig:results}d); and Ci (high) clouds more globally widespread (Fig.~\ref{fig:results}b). The highest occurrence of Ci clouds appears roughly over the inter-tropical convergence zone and is spatially correlated with Dc clouds, in agreement with \citet{Mace}. Most interestingly, the heatmaps show great spatial similarities with the ones reported by \cite{sassen_2008_1,sassen_2008_2,sassen_2009}, where the authors studied occurrences of cloud classes labeled by CloudSat over a period of one~ \cite{sassen_2008_1,sassen_2008_2} and two years \cite{sassen_2009}. All occurrences are shown in Fig.~\ref{fig:occurrences}.

\begin{figure}[h!]
\centering
\subfigure[Ci clouds (LWP).]
  {
  \includegraphics[width=.20\textwidth]{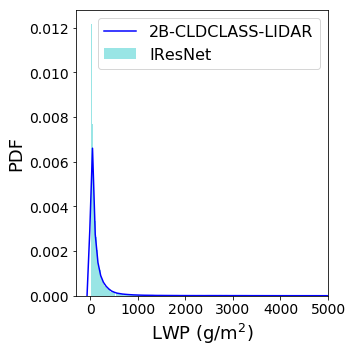}
  }
\subfigure[Ci clouds (COT).]
  {
  \includegraphics[width=.20\textwidth]{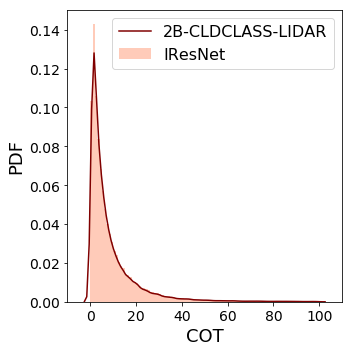}
  }  
\subfigure[As clouds (LWP).]
  {
  \includegraphics[width=.20\textwidth]{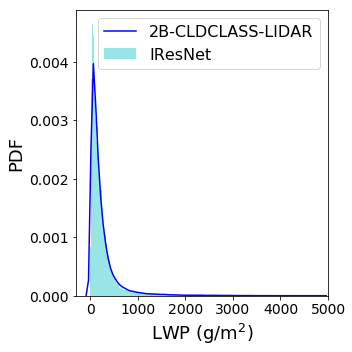}
  }
\subfigure[As clouds (COT).]
  {
  \includegraphics[width=.20\textwidth]{figures/COT-IResNet-class-0-logWeights.png}
  }  
\subfigure[Ac clouds (LWP).]
  {
  \includegraphics[width=.20\textwidth]{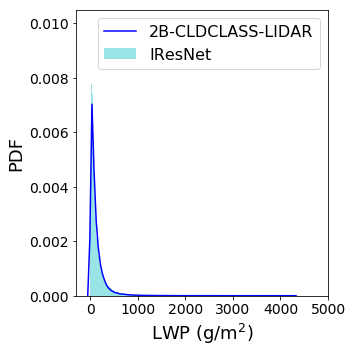}
  }
\subfigure[Ac clouds (COT).]
  {
  \includegraphics[width=.20\textwidth]{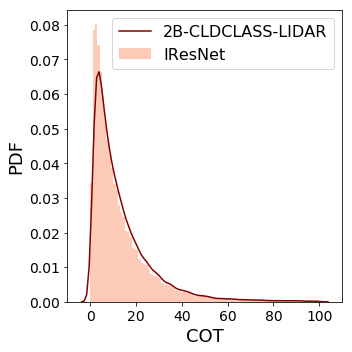}
  }  
\subfigure[St clouds (LWP).]
  {
  \includegraphics[width=.20\textwidth]{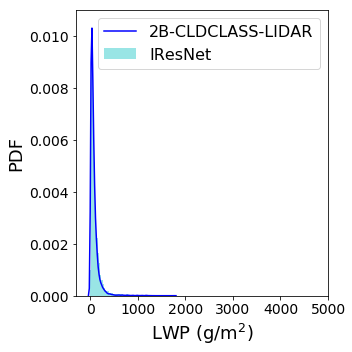}
  }
\subfigure[St clouds (COT).]
  {
  \includegraphics[width=.20\textwidth]{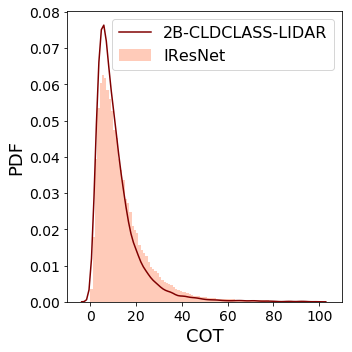}
  }  
\subfigure[Sc clouds (LWP).]
  {
  \includegraphics[width=.20\textwidth]{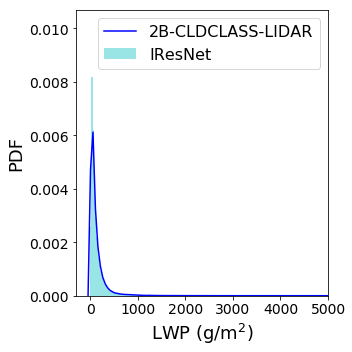}
  }
\subfigure[Sc clouds (COT).]
  {
  \includegraphics[width=.20\textwidth]{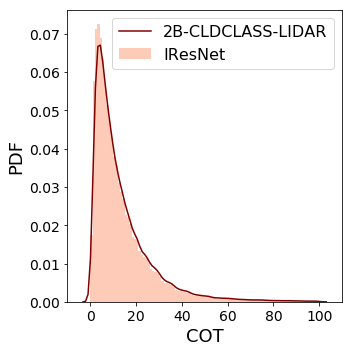}
  }  
\subfigure[Cu clouds (LWP).]
  {
  \includegraphics[width=.20\textwidth]{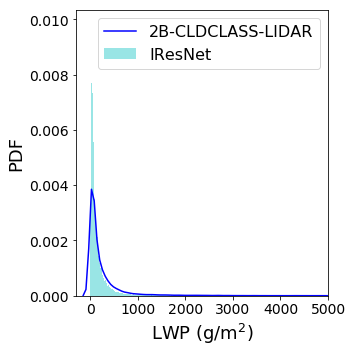}
  }
\subfigure[Cu clouds (COT).]
  {
  \includegraphics[width=.20\textwidth]{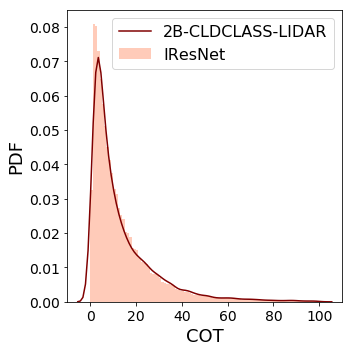}
  }  
\subfigure[Ns clouds (LWP).]
  {
  \includegraphics[width=.20\textwidth]{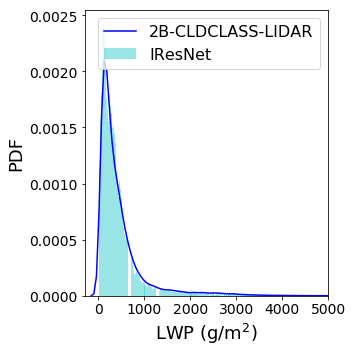}
  }
\subfigure[Ns clouds (COT).]
  {
  \includegraphics[width=.20\textwidth]{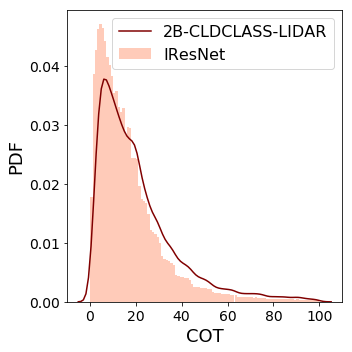}
  }  
\subfigure[Dc clouds (LWP).]
  {
  \includegraphics[width=.20\textwidth]{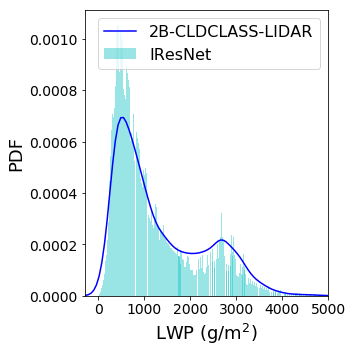}
  }
\subfigure[Dc clouds (COT).]
  {
  \includegraphics[width=.20\textwidth]{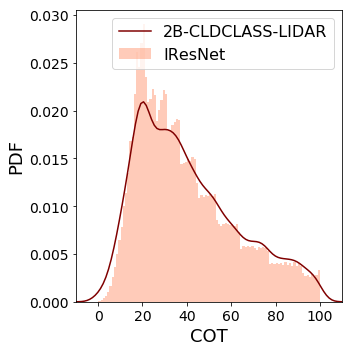}
  }  
\caption{PDFs of the liquid water path (LWP) and cloud optical thickness (COT) for the cloud classes predicted by the IResNet for January 2008. LWP(COT) histograms are drawn in blue(red).}
\label{fig:PDFs}
\end{figure}

As an additional physical-based evaluation we consider the distributions of the liquid water path (LWP) and cloud optical thickness (COT) variables for the predicted classes and the ground truth given by CloudSat~\footnote{Note that both LWP and COT features were not used for training.}. LWP defines the total amount of liquid water present in the whole atmospheric column on a given point. COT is a measure of the thickness between the bottom and top of a cloud. 
Qualitatively, differences between the predicted distributions and the ground truth are minimal for both variables. Results are shown in Figure~\ref{fig:PDFs}.
In Table~\ref{tab:perfs-Physics} we also provide a quantitative comparison between predicted distributions and CloudSat ones by means of both the Kullback–-Leibler divergence and the Wasserstein distance. 
We note that the class "Deep Convection" is the one associated with largest accuracy ($98.84\%$)  
despite having relatively large differences with the ground truth values in both the LWP and COT variables (see Table~\ref{tab:perfs-Physics}).

In general, as the available supervision is minimal, we argue that it is inadequate to gauge the quality of a method considering exclusively its accuracy on testing samples. Therefore, a physical-based evaluation is more tailored to cloud classification studies, it does not suffer from the minimal supervision and it should always complement the more basic metrics discussed in Table~\ref{tab:perfs}. 


\begin{table}
  \caption{Machine-Learning-based validation: IResNet classification results on test set.}
  \label{tab:perfs}
  \centering
  \begin{scriptsize}
  \begin{tabular}{lccccccccc}
    \toprule
    & Ci & As & Ac & St & Sc & Cu & Ns & Dc & Mean \\
    \midrule
    Accuracy (\%) & 81.30 & 84.50 & 88.29 & 97.73 & 88.90 & 92.40 & 90.92 & 98.84 & 90.36     \\
    F1 score & 0.68 & 0.43 & 0.45 & 0.58 & 0.80 & 0.40 & 0.47 & 0.58 & 0.55       \\
    IoU index     & 0.52 & 0.28 & 0.29 & 0.41 & 0.66 & 0.25 & 0.31 & 0.41 & 0.39  \\ 
    \bottomrule
  \end{tabular}
  \end{scriptsize}
\end{table}

\begin{table}
  \caption{Physical-based validation: IResNet classification results for liquid water path (LWP) and cloud optical thickess (COT) distributions. We compare the distributions predicted by the IResNet with the ground truths using both Kullback-Leibler (KL) divergence and the Wasserstein (W) distance.}
  \label{tab:perfs-Physics}
  \centering
  \begin{scriptsize}
  \begin{tabular}{llccccccccc}
    \toprule
     & metric & Ci & As & Ac & St & Sc & Cu & Ns & Dc \\
    \midrule
    LWP & KL div. & $0.11\cdot 10^{-1}$ & $0.38 \cdot 10^{-2}$ & $0.71 \cdot 10^{-2}$ & $0.32 \cdot 10^{-1}$ & $0.33 \cdot 10^{-2}$ & $0.39 \cdot 10^{-1}$ & $0.23 \cdot 10^{-1}$ & $0.35 \cdot 10^{-1}$ \\
    & W dist. & $0.12\cdot 10^{-3}$ & $0.54\cdot 10^{-4}$ & $0.76\cdot 10^{-4}$ & $0.18\cdot 10^{-3}$ & $0.51\cdot 10^{-4}$ & $0.21 \cdot 10^{-3}$ & $0.95 \cdot 10^{-4}$ & $0.15 \cdot 10^{-3}$ \\
    COT & KL div. & $0.47\cdot 10^{-2}$ & $0.14 \cdot 10^{-1}$ & $0.34 \cdot 10^{-1}$ & $0.34 \cdot 10^{-1}$ & $0.13 \cdot 10^{-1}$ & $0.28 \cdot 10^{-1}$ & $0.57 \cdot 10^{-1}$ & $0.47 \cdot 10^{-1}$ \\
    & W dist. & $0.46\cdot 10^{-4}$ & $0.58\cdot 10^{-4}$ & $0.68\cdot 10^{-4}$ & $0.21\cdot 10^{-3}$ & $0.38\cdot 10^{-4}$ & $0.99 \cdot 10^{-4}$ & $0.17 \cdot 10^{-3}$ & $0.73 \cdot 10^{-4}$ \\
    \bottomrule
  \end{tabular}
  \end{scriptsize}
\end{table}

\subsection{Conclusions}
In this work we first proposed \ourDataset, a new benchmark dataset for training and evaluating global cloud classification models.  It consists of one year of 1km resolution MODIS hyperspectral images merged with pixel-width tracks of CloudSat labels. We think that this is an important step for engaging the Machine-Learning community to develop innovative methods and solutions to climate related problems. In particular, the proposed dataset presents several important challenges: (a) labels make up for less than $1\%$ of the data (weakly-labelled data), (b) a pixel can have multiple labels (multi-labelled data), (c) certain type of clouds (i.e., deep convection) are underrepresented (class imbalance).
Moreover, within a single cloud class we can still distinguish a rich variety of cloud organizations at the mesoscale (from 5 to several hundred kilometers) and slightly different physical properties, implying the existence of sub-classes for each given class. Proposing novel unsupervised models that directly discover fine-grained classes with only access to the observed coarse labels could be an important new line of research for both the Climate and Machine-Learning community.

We also provided a first high resolution spatiotemporal cloud classification baseline performance on \ourDataset. 
To complement the standard ML prediction scores, we made use of the validation channels of \ourDataset~to analyse the results from a physical perspective. First, the occurrences over the analyzed month are found to be qualitatively similar to previous studies. Second, we found that our baseline results are physically reasonable in terms of the liquid water path and cloud optical thickness distributions of the predicted classes.  
The reported analysis is quantitative and physical, but limited to January 2008. 
Since CloudSat needs 16 days to complete a cycle, we leave a rigorous comparison of the (predicted) monthly occurrences for future studies, using 1 year of classification.

\clearpage

\subsubsection*{Data availability}
\ourDataset~is publicly available at \url{https://www.dropbox.com/sh/i3s9q2v2jjyk2it/AACxXnXfMF5wuIqLXqH4NJOra?dl=0}.
The code used for extracting \ourDataset~is hosted at~\url{https://github.com/FrontierDevelopmentLab/CUMULO}.

\subsubsection*{Acknowledgments}

This work is the result of the 2019 ESA Frontier Development Lab~\footnote{\url{https://fdleurope.org/}} (FDL) Atmospheric Phenomena and Climate Variability challenge. We are grateful to all organisers, mentors and sponsors for providing us this opportunity. We thank Google Cloud for providing computing and storage resources to complete this work. Finally, we thank Yarin Gal for helpful discussions and Sylvester Kaczmarek for his help and support in coordinating the work.

\bibliographystyle{unsrtnat}
\bibliography{NeurIPS-2019}

\end{document}